# Experimental Evidence about "A factorisation algorithm in adiabatic quantum computation" by T. D. Kieu

Richard H. Warren – Lockheed Martin Corporation, Retired

rhw3@psu.edu

Abstract. Computations show that the logic about a quantum factoring algorithm does not hold in reality on a D-Wave quantum computer. We demonstrate this for the integers 15 = 3 x 5, 91 = 7 x 13 and 899 = 29 x 31. The likely cause is the D-Wave hardware that does not accept input terms that are a number, i.e., only terms that contain a Boolean variable can be an input. Without terms that are numbers, the relative magnitude of the coefficients in the factoring algorithm is too great to differentiate values.

1. <u>Introduction</u>. Factoring an integer is a useful topic due to its important application in the cryptanalysis field. Several public-key schemes, such as RSA, are based on the current status that factoring arbitrary integers with more than 1000 digits is not feasible. Currently there is no polynomial time algorithm for factoring on a classical computer. Factoring on a quantum computer offers a potential speedup over factoring on a classical computer.

Expression (1) in Kieu's paper [1] is a minimization function to factor a positive integer $N$ into two factors $x$ and $y$. The function is

$$N^2(N - xy)^2 + x(x - y)^2 \qquad (1)$$

Kiev has noted that the logic of minimizing (1) in adiabatic quantum computation indicates high probability of finding factors $x$ and $y$ such that $N = xy$, $x \leq \sqrt{N} \leq y$ and that $x = 1$ if and only if $N$ is a prime number. However, in reality there are difficulties that prevent this. A major problem seems to be omission of constants in the D-Wave hardware for adiabatic computation which results in very large coefficients on the left side of (1) and small coefficients on the right side of (1). The small coefficients are lost in the size of the large ones.

The integers 15 = 3 x 5, 91 = 7 x 13 and 899 = 29 x 31 could not be factored by Kieu's method. No other integers were attempted. In contrast, the authors of [2] claim that its algorithm can be used to quantum factor all semiprimes up to 200,099 and presents substantiation for several of these semiprimes in their Table 2.

We used a 1,000 qubit D-Wave quantum computer and the D-Wave software dw which is a utility tool that provides shell access to many common application functions for quantum computing on an annealing processor. These functions include: forming the Hamiltonian matrix, embedding the Boolean variables in the qubits, reducing large coefficients by distributing their value over several qubits, and flagging solutions that satisfy assertions. This is a tremendous boost to users by freeing them to concentrate on the problem to be solved.

2. <u>Analysis</u>.  When substituting Boolean expressions for $x$ and $y$ in (1) and expanding, we noted that the terms $N^2 - 2N^3 + N^4$ have no role in the D-Wave minimization objective which is (2) in [3].  When these terms are omitted from (1), the result is very negative and all coefficients are divisible by 4.  Therefore, the minimization function that was executed on a D-Wave quantum computer is

$$[N^2(N - xy)^2 - N^2 + 2N^3 - N^4 + x(x - y)^2]/4 \qquad (2)$$

We simplified the implementation of (2) by requiring $x$ and $y$ to be odd integers, which means the smallest bit of $x$ and $y$ is 1.  We set $x = 1 + 2*x1 + 4*x2 + 8*x3 + 16*x4$ and $y = 1 + 2*y1 + 4*y2 + 8*y3 + 16*y4$ where xi and yi are Boolean variables for i = 1, 2, 3, 4.

Table 1.  Indication of the approximate numbers to implement (2) for $N = xy$.  D-Wave hardware values are imprecise due to scaling, noise, inability to distinguish between the ground state and an excited state, and the analog manner in which the hardware represents numbers.

| $N$ | Factor $x$ | Factor $y$ | $N^2(N - xy)^2$ | $-N^2 + 2N^3 - N^4$ | $x(x - y)^2$ | Sum for (2) |
|---|---|---|---|---|---|---|
| 15 | 3 | 5 | 0 | -44,100 | 12 | -11,022 |
| 91 | 7 | 13 | 0 | -67,076,100 | 252 | -16,768,962 |
| 899 | 29 | 31 | 0 | -6.5173651e11 | 116 | -1.6293412e11 |

The leading term $N^2$ has no essential role in [1].  If it is deleted, then $2N^3 - N^4$ are eliminated from (2) and the computations are simplified.  The leading $x$ in $x(x - y)^2$ causes $x \leq \sqrt{N}$; but $x \geq \sqrt{N}$ is an equally valid factor.  Thus, (1) can be simplified by deleting the leading $x$.

The factor $x = 1$ is not wanted unless $N$ is a prime.  It is discouraged by $(x - y)^2$ which maximizes at $x = 1$ and $y = N$.

3. <u>Implementation of (2)</u>.  The D-Wave software dw was used on the 1000 qubit computer DW2X_LANL_1 at Los Alamos National Laboratory.  A successful factoring occurs when a solution is tagged "valid" which means all chains are intact and all assert statements are true.  Our only assert statement is $N - xy = 0$.  A chain is a string of connected qubits that represent a logical qubit.  A chain can be used to break a large number into a sum of smaller numbers.  The software dw automates this.

Ancillary variables are introduced to reduce terms whose degree is greater than 2.  Our success criteria for factoring does not require all ancillary variables to have correct representation in an execution.

Two parameters are set in a bind statement.  The parameter param_chain weights chains to ensure that they do not break.  The parameter S weights the expressions for ancillary variables to ensure their representations are interpreted accurately.

An execution is not repeatable because the embedding of the variables in the qubits is non-deterministic.  A solution is "valid" if the assert statement is true and all chains are intact.

All executions of the software asked for 1,000 samples. All results were Distinct Samples = 1,000 and Total Valid = 0. Samples are distinct if they have different parts of chains intact or have different factorings. A valid sample has all chains unbroken and a correct factoring.

For $N$ = 15, we executed the software for param_chain between 10 & 100 in steps of 10, between 100 & 300 in steps of 20, between 300 & 11,400 in steps of 100; and S = param_chain / 3. During the incrementing we tried many other values for S.

For $N$ = 91, we executed the software for 300 ≤ param_chain ≤ 11,400 in steps of 150 and S = param_chain / 3. While incrementing we tried other values for S.

For $N$ = 899, we executed the software for 300 ≤ param_chain ≤ 9,900 in steps of 300 and S = param_chain / 3. At some steps we tried other values for S.

We could not factor the integers 15, 91 and 899. The number of qubits used varied and were less than 200 for all executions.

Two careful checks were made of the factoring routine. One of the checks used a new execution of Mathematica for the terms. The factoring routine is available upon request and is strongly commented.

4. <u>Deductions</u>. We have shown that Kieu's algorithm cannot factor 15, 91 and 899. We conclude that it will not factor most odd integers in the range 15 to 899.

Possible reasons that Kieu's algorithm fails to factor: 1) Table 1 shows that the size of the numbers for $N^2(N-xy)^2 - N^2 + 2N^3 - N^4$ is huge compared to the size for $x(x-y)^2$ in the objective function (2) on D-Wave hardware. 2) the scale factor for $N$ = 899 varies around 2.63e-10 which seems to prohibit distinguishing small variations in the objective function (2). 3) omission of constants by the D-Wave hardware is a likely cause for reasons 1 and 2.

Two routines have factored successfully using the software dw. Therefore, the software dw is not a suspect of causing the failure of function (2) to factor.

Preliminary work with $x$ sized to 3 bits and $y$ sized to 4 bits factored 15 and 35 with param_chain = 450 and S = 150. Therefore, the range of the values that were checked for these two parameters is appropriate when the bit size of $x$ and $y$ is increased to 5.

5. <u>Conclusion</u>. Our results indicate that the relative size of the large and small coefficients in expressions (1) and (2) is not suitable for quantum factoring on the current D-Wave hardware.

Acknowledgements. Tien Kieu approved the experiment and checked the input. Denny Dahl assisted with the software tool dw, provided technical assistance, and reviewed a draft of this report. Scott Pakin asked questions that improved the content of this paper. Los Alamos National Laboratory granted access to their D-Wave quantum computer. Jesus Christ said "without Me you can do nothing."

References

1. T. D. Kieu (2018), A factorisation algorithm in adiabatic quantum computation, arXiv:1808.02781v2.
2. R. Dridi and H. Alghassi (2017), Prime factorization using quantum annealing and computational algebraic geometry, *Sci. Rep.*, Vol. 7, 43048.  Erratum: *Sci. Rep.*, Vol. 7, 44963.
3. R. H. Warren (2018), Mathematical methods for a quantum annealing computer, Journal of Advances in Applied Mathematics, Vol. 3, 82-90; www.isaacpub.org/images/PaperPDF/JAAM_100085_2018061111161573385.pdf.